\documentclass[sigconf]{acmart}
\usepackage{subcaption}
\usepackage{amsmath}
\usepackage{multirow}
\AtBeginDocument{%
  \providecommand\BibTeX{{%
    \normalfont B\kern-0.5em{\scshape i\kern-0.25em b}\kern-0.8em\TeX}}}

\setcopyright{acmlicensed}
\copyrightyear{2025}
\acmYear{2025}
\acmDOI{XXXXXXX.XXXXXXX}

\acmConference[C'25]{Conference}{Nov 2025}{Australia}

\acmISBN{978-1-4503-XXXX-X/18/06}

\begin{document}

\title{Double-Signed Fragmented DNSSEC for Countering Quantum Threat}


\author{Syed W. Shah, Lei Pan}
\affiliation{%
  \institution{Deakin Cyber Research and Innovation Centre, Australia \\
  Cyber Security Cooperative Research Centre, Australia}
  \streetaddress{}
  \city{}
  \country{}}
  \email{syed.shah, l.pan@deakin.edu.au}

\author{Dinh Duc Nha Nguyen, Robin Doss}
\affiliation{%
  \institution{Deakin Cyber Research and Innovation Centre, Australia \\
  Cyber Security Cooperative Research Centre, Australia}
  \streetaddress{}
  \city{}
  \country{}}
  \email{dinh.nguyen, robin.doss@deakin.edu.au}

\author{Warren Armstrong}
\affiliation{%
  \institution{Quintessence Labs}
  \streetaddress{}
  \city{Canberra}
  \country{Australia}}
\email{wa@quintessencelabs.com}

\author{Praveen Gauravaram}
\affiliation{%
  \institution{Tata Consultancy Services (TCS)}
  \city{Brisbane}
  \country{Australia}}
  \email{p.gauravaram@tcs.com}



\begin{abstract}
DNSSEC, a DNS security extension, is essential to accurately translating domain names to IP addresses. Digital signatures provide the foundation for this reliable translation; however, the evolution of \emph{Quantum Computers} has made traditional digital signatures vulnerable. In light of this, NIST has recently selected potential \emph{post-quantum} digital signatures that can operate on conventional computers and resist attacks made with \emph{Quantum Computers}. Since these \emph{post-quantum} digital signatures are still in their early stages of development, replacing pre-quantum digital signature schemes in DNSSEC with post-quantum candidates is risky until the post-quantum candidates have undergone a thorough security analysis. Given this, herein, we investigate the viability of employing \emph{Double-Signatures} in DNSSEC, combining a post-quantum digital signature and a classic one \footnote{A preliminary version of this article was accepted in ITNAC 2025}. The rationale is that double-signatures will offer protection against quantum threats on conventional signature schemes as well as unknown non-quantum attacks on post-quantum signature schemes, hence even if one fails, the other provides security guarantees. However, the inclusion of two signatures in the DNSSEC response message doesn't bode well with the maximum allowed size of DNSSEC responses (i.e., 1232B, a limitation enforced by the MTU of physical links). To counter this issue, we leverage a way to do application-layer fragmentation of DNSSEC responses with two signatures. We implement our solution on top of OQS-BIND and, through experiments, show that the addition of two signatures in DNSSEC and application-layer fragmentation of all relevant resource records and their reassembly does not have a substantial impact on the efficiency of the resolution process and thus is suitable for the interim period at least until the quantum computers are fully realized.
\end{abstract}


\maketitle
\section{Introduction}
\label{intro}
The Domain Name System (DNS) is responsible for translating human-readable domain names into machine-understandable IP addresses. As can be imagined, any vulnerability in this system can potentially make such translation precarious, leading users to malicious servers instead of their intended destinations. The Security Extension of DNS - i.e., Domain Name System Security Extensions (DNSSEC) helps establish that the messages received by a client are from an authorized DNS server and have not been altered in transit. DNSSEC primarily leverages \emph{digital signatures} for establishing the aforementioned properties of \emph{authenticity} and \emph{integrity}. However, the underlying security assumptions of conventional digital signature algorithms - i.e., the computational difficulty of integer factorization and discrete logarithms will not hold when an attacker has access to a Cryptographically Relevant Quantum Computer (CRQC) \cite{CRQC,shor1994algorithms}. To mitigate this problem, the National Institute of Standards and Technology (NIST) has recently selected three post-quantum digital signatures - i.e., FALCON \cite{Falcon}, CRYSTALS-DILITHIUM \cite{Dilithium}, and SPHINCS+ \cite{SPHINCS}. An important envisioned property of these post-quantum digital signature algorithms is that they can run on classical computers but also withstand the attacks conducted using CRQCs, thereby offering sufficient protection against the impending threat of CRQCs. For DNSSEC to continue offering the needed \emph{authenticity} and \emph{integrity} in the post-quantum era, it must be updated with post-quantum digital signatures. However, for successful transitioning to the post-quantum era, the \emph{Interim Period} - i.e., from now until CRQCs are fully realized or until the security of post-quantum candidates is fully established- presents a particular challenge. Precisely, in preparation for the post-quantum era, simply switching pre-quantum cryptography (such as conventional digital signatures in DNSSEC) with post-quantum cryptography (such as the post-quantum candidates selected by NIST) in the interim period may lower overall security \cite{ENSIA_Report}. The reason is that the selected post-quantum candidates (both key encapsulation mechanisms and digital signatures) have not yet gone through a thorough cryptanalysis against both classical and quantum attacks. The selected post-quantum candidates may be compromised even by conventional classical computers in the near future. Therefore, there is an inherent risk in simply switching from pre-quantum cryptography to post-quantum cryptography alone in that overall security may be lowered not only against CRQCs but also against today's classical computers (see further details in Section \ref{quantum_threat}).

Aligned with the aforementioned discussion and the recommendations from the European Union Agency for Cybersecurity (ENISA), we in this paper investigate the plausibility of \emph{Double-Signed DNSSEC} by combining the pre-quantum and post-quantum digital signatures for the interim period. However, the size of signatures and public keys of post-quantum candidates selected by NIST is very large (i.e., selected post-quantum signatures are 11\(\times\) to 122\(\times\) and public keys are 14\(\times\) to 20\(\times\) of their pre-quantum counterparts). These sizes further increase when double signatures are incorporated into the DNSSEC resolution process to ensure authenticity and integrity. This increased size of signatures and public keys significantly impacts the DNSSEC resolution process \cite{dnssec_size_issue}. Precisely, since UDP is the preferred transport protocol for DNSSEC because of it being lightweight, any response message that exceeds 1232B triggers IP-level fragmentation, which is unreliable. The maximum size of DNSSEC response messages - i.e., 1232B is obtained as: (IPv6 MTU=1280)-(IPv6 Header=40)-(UDP Header=8). With double signatures, the sizes of DNSSEC response messages that carry different resource records for facilitating reliable address resolutions are much larger than the threshold size of 1232B. Therefore, these large DNSSEC messages necessitate IP fragmentation which is unreliable over UDP (i.e., some fragments may never reach the resolver) \cite{Frag_issue}. Due to this issue, the standard DNSSEC offers a fallback mechanism via TCP for sending large response messages without resorting to fragmentation. Precisely, when the response size exceeds the threshold (i.e., the resolver's advertised EDNS(0) buffer size), name servers send back a truncated response (i.e., by setting the TC flag), indicating the resolver to retry via TCP (and thus negatively contributing to the address resolution time). However, many name servers lack support for TCP in addition to high-resolution time with TCP due to it not being as lightweight as UDP \cite{Frag_issue, TCP-issue, resolver_issue}.

To circumvent the aforementioned issues, a potential alternative is to perform the fragmentation at the application layer (i.e., within the DNS protocol itself). While there exist ways to tackle the application layer fragmentation (see \cite{goertzen2022postquantum, frag_new}), none of them can accommodate double signatures and public keys. Therefore, to accomplish this, first, we constructed a Docker-based testbed on an Amazon EC2 instance using commercial-grade DNS software - i.e., \emph{BIND9} (OQS-BIND, see full details on our testbed in Section \ref{setup}). Second, we explored generating two signatures (pre-quantum and post-quantum) on the \emph{Name Server} using the default BIND9 tools. Third, we devised a way to fragment the DNSSEC responses with two signatures and public keys on the name servers and accurately reassemble them on the resolver. We also modified the source code of the BIND9 resolver to enable the verification of both signatures before marking the response as authenticated. Finally, using our testbed, we conducted an empirical analysis to show that our implementation can handle two signatures in DNSSEC simultaneously and quantify the impact of double signatures on the DNSSEC resolution process. Our experiments show that double signatures have a negligible impact on the average resolution time of DNSSEC compared to only post-quantum digital signatures. Therefore, this analysis confirms the efficacy of double signatures in DNSSEC for the interim period until CRQCs are fully realized. Our main contributions are summarized as follows:
\begin{itemize}
    \item We developed a fully functional Docker-based DNSSEC testbed over an Amazon EC2 instance using the widely used BIND9 software that can handle both pre-quantum and post-quantum signatures and public keys simultaneously over UDP in a single DNSSEC response message.
    \item We thoroughly investigated the source code for BIND9 and modified the resolver component to enable the re-assembly and verifications of both pre-quantum and post-quantum digital signatures simultaneously.
    \item Through empirical analysis, we show that the addition of double signatures in the DNSSEC resolution process did not affect the performance. Hence, it is recommended for the interim period until CRQCs are fully realized.
\end{itemize}

The rest of the paper is organized as follows. Section \ref{dns} presents a brief overview of DNS, while Section \ref{quantum_threat} discusses the overall quantum threat and the motivation for using double signatures in DNSSEC. Section \ref{size_issue} presents the details of the maximum message size issue with double signatures and how to handle them in the resolution process. Section \ref{setup} details our test setup and experimental methodology. Results are discussed in Section \ref{result} and related works and concluding remarks are contained in Sections \ref{related_work} and \ref{conclusion}, respectively.

\section{Domain Name System (DNS)}
\label{dns}
The Domain Name System (DNS) is a distributed database that plays a pivotal role in the functioning of today's Internet by facilitating the translation of human-readable domain names into corresponding machine-understandable IP addresses. In general, DNS is divided into various \emph{zones} that correspond to a particular level of granularity in the aforementioned translation process. These \emph{zones} contain various \emph{types} of \emph{resource record} (RR) that correspond to \emph{labels}. These \emph{resource records} are generally used to look up the IP addresses of domain names, name servers of various zones, and various other types of data needed for address translation.

For an abstract-level understanding of the DNS translation process, imagine a client who queries the IP address of a domain name \emph{example.com}. For this translation, the client will first send a query to a \emph{(caching) resolver} which will handle all the subsequent translation processes on the client's behalf.
Assuming that the resolver has not previously cached this answer (i.e., the IP address of \emph{example.com}), it will query the \emph{root server} for the name server of \emph{.com} domain names. Once the resolver receives the response, it queries this name server for the name server(s) of \emph{example.com}. Finally, once the resolver knows the corresponding name server, it queries it for the IP address of the domain name \emph{example.com} and returns it to the client.

\begin{figure}[!h]
\centering
\includegraphics[width=0.5\textwidth]{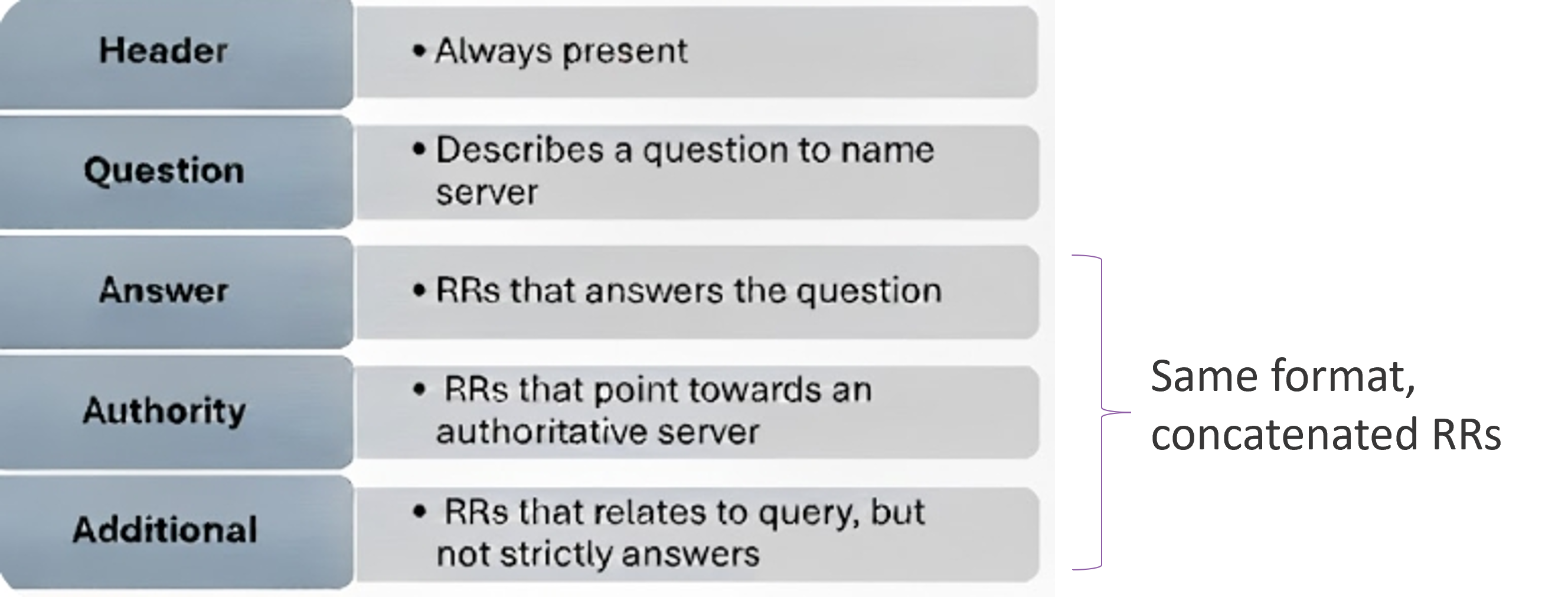}
\caption{Generic DNS Message Format}
\label{Message_Format}
\end{figure}

\begin{figure}[!h]
\centering
\includegraphics[width=0.48\textwidth]{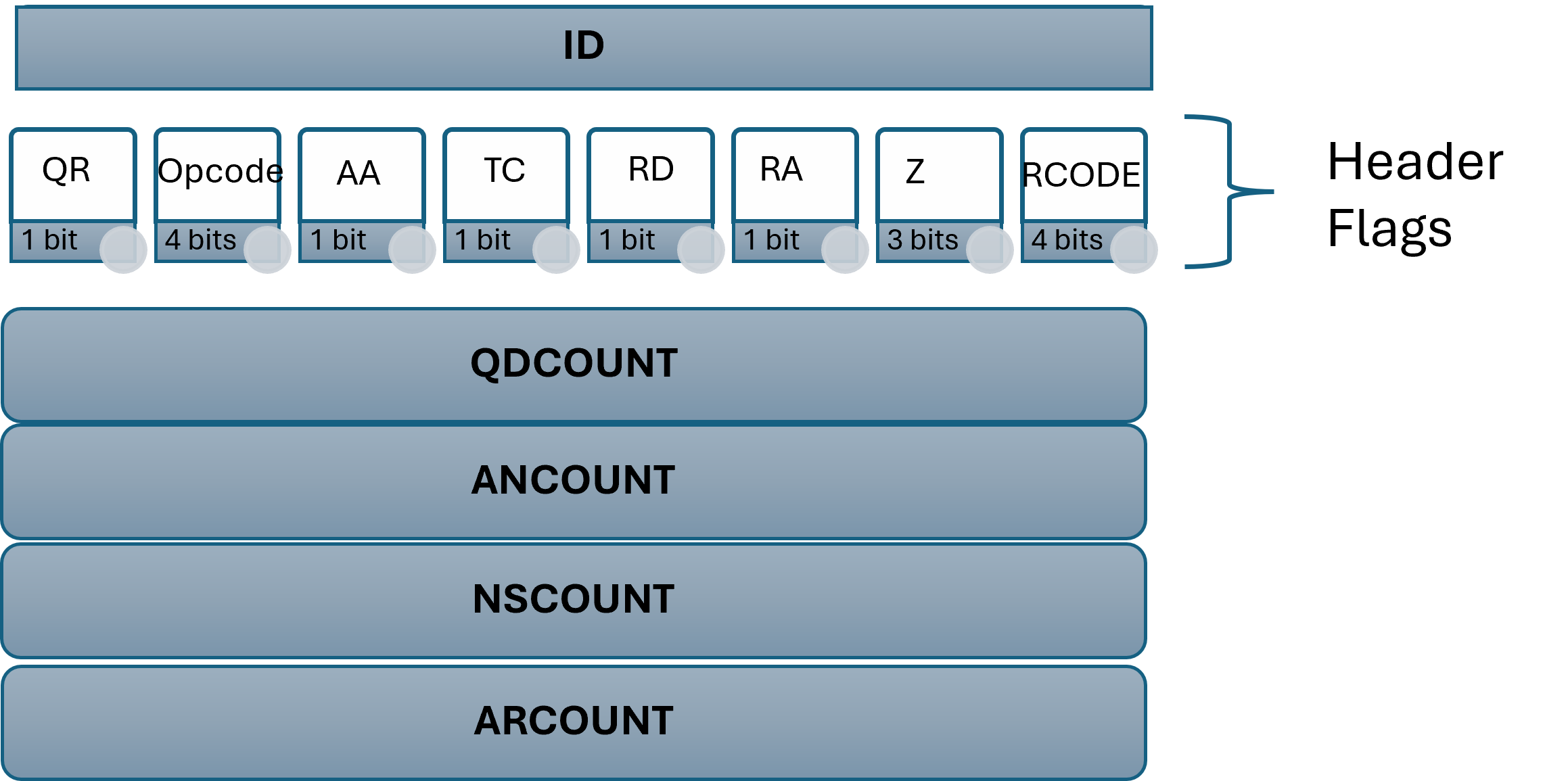}
\caption{DNS Header}
\label{Header_Format}
\end{figure}

As specified in RFC 1035, the generic DNS messages exchanged between the resolver and name servers have five sections \cite{RFC1035}. Figure \ref{Message_Format} shows an overview of a generic DNS message. The \emph{header} section is always present and its structure is depicted in Figure \ref{Header_Format}. The \emph{Question} section encapsulates the resolver's query whose IP address needs to be resolved. This section has three main fields - i.e., \emph{QName} (contains encoded query), \emph{QType} (mentions the type of needed resource recorded in response), and \emph{QClass} (contains the class of query, mostly specifies IN). As shown in Figure \ref{Message_Format}, the last three sections have a similar format - i.e., concatenation of various RRs. The RRs are actual data entries in the DNS database that provide information related to the queried domain name and help the actual translation of domain names to corresponding IP addresses. The top-level format of RRs is shown below in Figure \ref{RR_Format}. Going back to our earlier example of \emph{example.com}, an RR with \emph{Name} `example.com' and \emph{TYPE} `A' will contain an IP address of 32 bits in the RDATA field, and with \emph{TYPE} `NS', the RDATA field will contain the domain name of the authoritative server for \emph{example.com}. The \emph{Answer} section in a response message contains RRs that answer the query from the resolver, while the \emph{Authority} section contains RRs that point to the authoritative server. The \emph{Additional} section contains RRs that somehow relate to the query but may not essentially contain the answers to the query.

\begin{figure}[!h]
\centering
\includegraphics[width=0.50\textwidth]{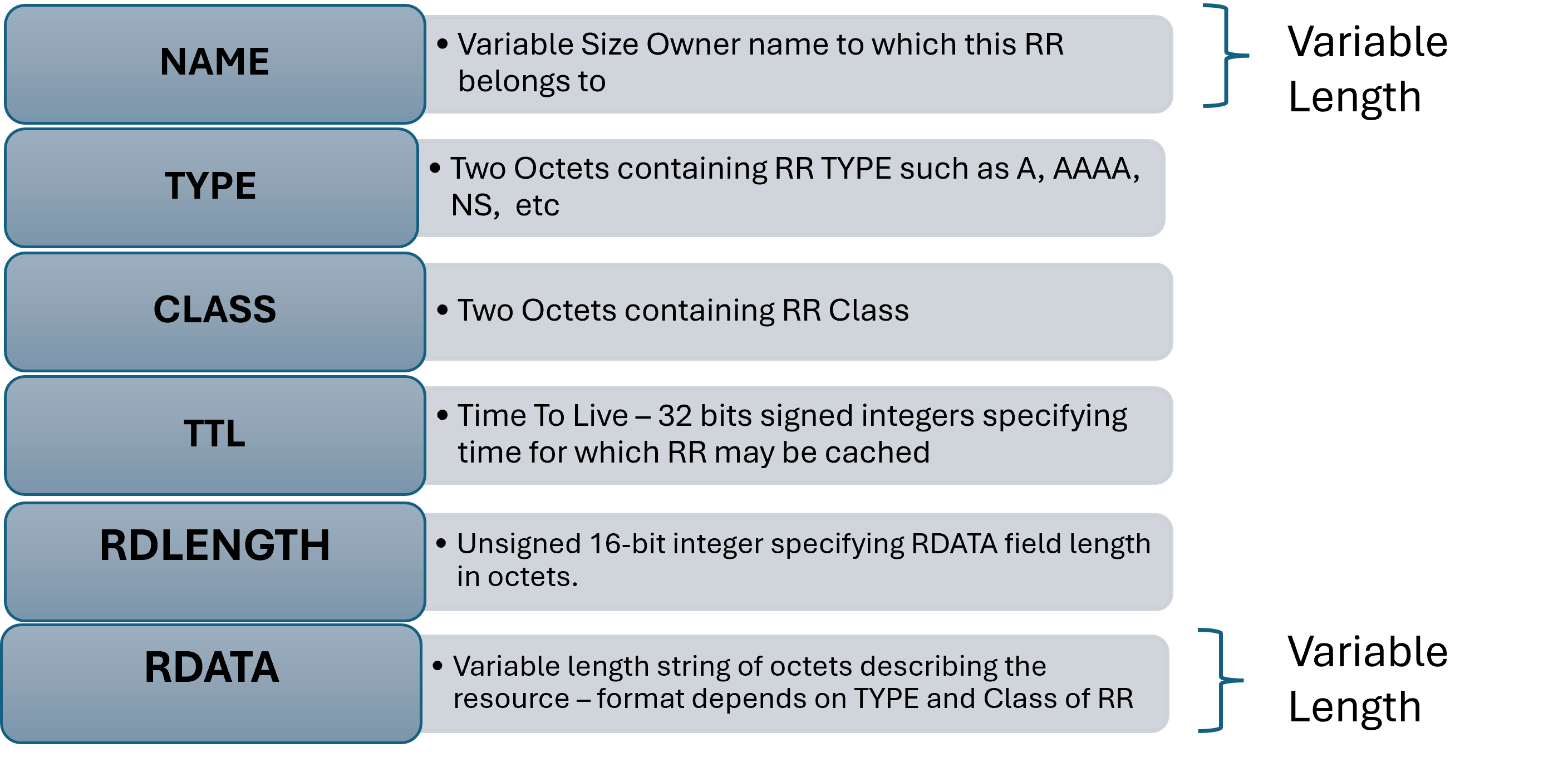}
\caption{RR's Top-Level Format}
\label{RR_Format}
\end{figure}
As discussed in Section \ref{intro}, DNSSEC leverages digital signatures to ensure source authenticity and data integrity in DNS. The resource record labels are not required to be unique in DNSSEC resource records of a particular type. The RRs of a specific type and label are grouped into a \emph{resource record set} (RRSet) that is signed using a specified digital signature algorithm. The corresponding digital signature is stored in \emph{RRSIG}, and the corresponding public key is published in \emph{DNSKEY}. In DNSSEC, for operational benefits, generally, two types of key pairs are generated - i.e., Zone Signing Keys (ZSK) and Key Signing Keys (KSK) (see details in RFC 6781 \cite{RFC6781}). The ZSK pair is used for signing and verifying the resource records, while the KSK pair is used for signing the ZSK and helps enable the \textit{chain of trust} in DNSSEC.

\begin{figure*}[!ht]
\centering
\includegraphics[width=0.93\textwidth]{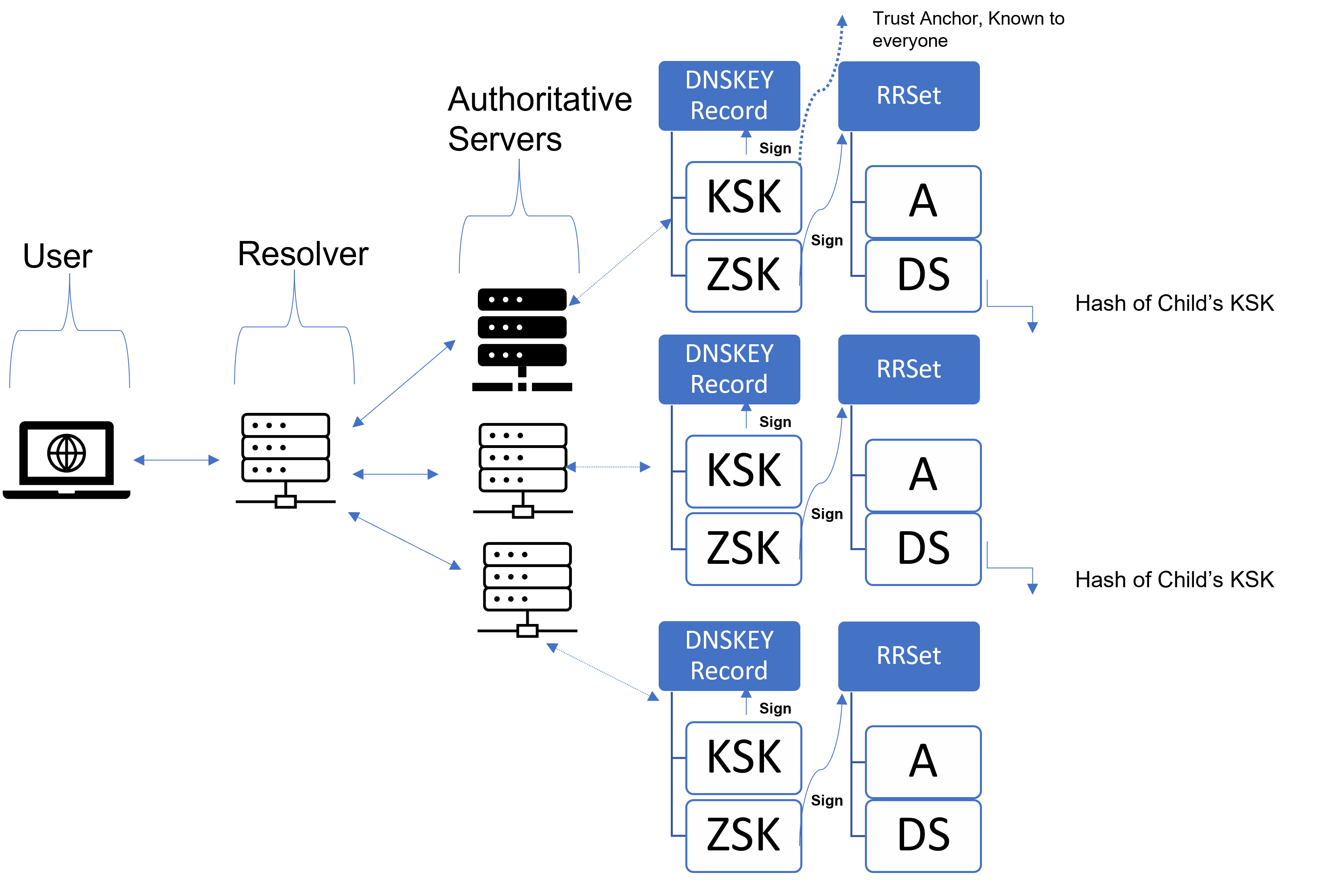}
\caption{Chain-of-Trust in DNSSEC}
\label{chainoftrust}
\end{figure*}

As explained above, the client initiates the translation process by querying the root server, then its children, and then its children's children, and finally reaching the name server that provides the queried IP address. This hierarchical process facilitates an important element of DNSSEC - i.e., the chain of trust that helps establish trust in public keys needed for signature verification. For the chain of trust, each zone that is queried as part of the resolution process must have a digest of public KSK kept as the \emph{Delegation Signer} resource record in its parent zone; otherwise, the public ZSK sent by the name server cannot be trusted. Since the root server(s) have no parent zone, it does not publish any \emph{delegation signer} (DS) resource record. Therefore, for the entire chain of trust, the public KSK of the root server is obtained out-of-band so that the root server's public KSK is pre-installed on most operating systems to eliminate the need for fetching and configuring the key(s). An overview of the chain of trust is shown in Figure~\ref{chainoftrust}.

\begin{figure}[!h]
\centering
\includegraphics[width=0.48\textwidth]{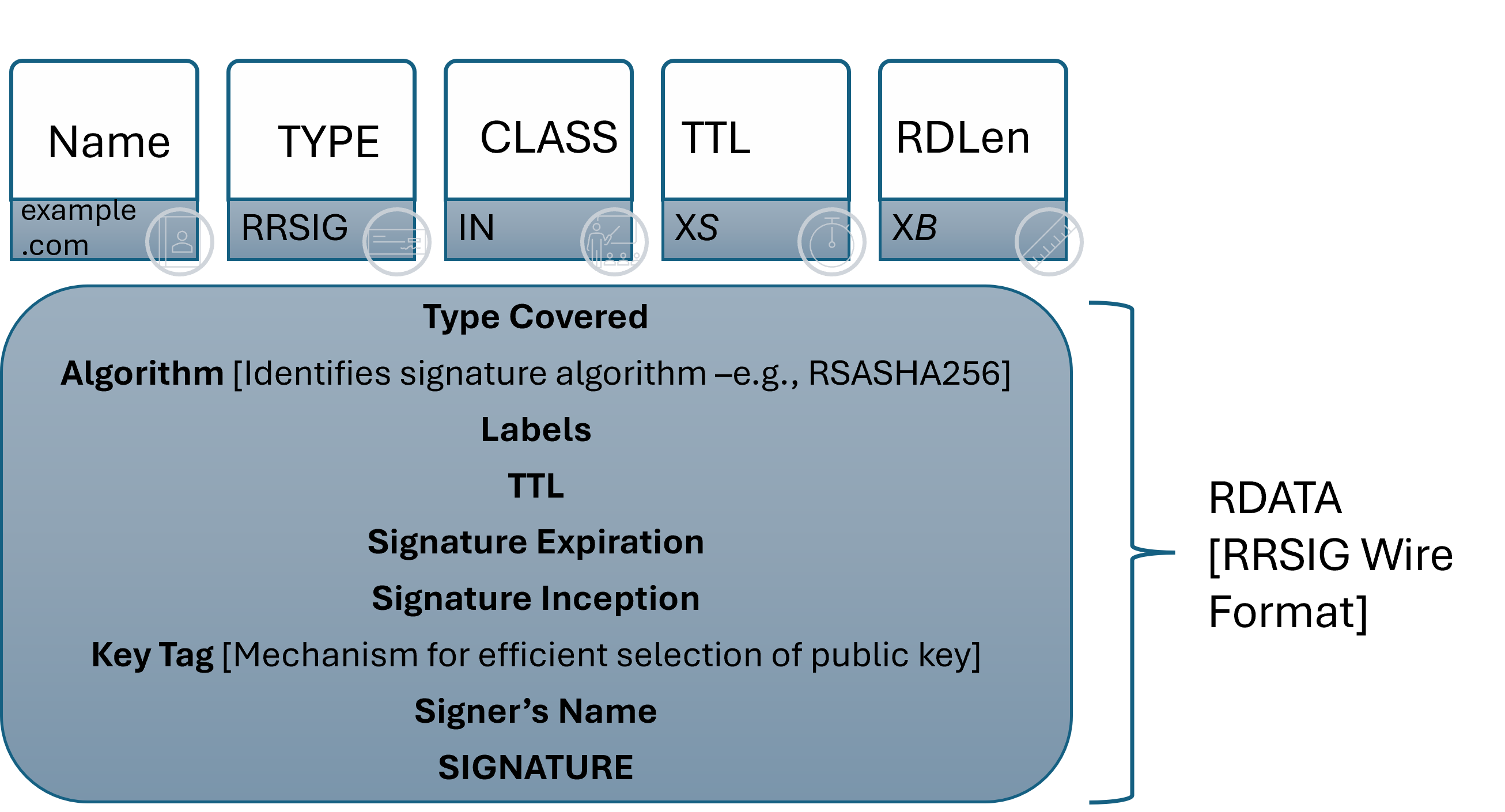}
\caption{RRSIG RR Wire Format }
\label{RRSIG_Format}
\end{figure}

\begin{figure}[!h]
\centering
\includegraphics[width=0.48\textwidth]{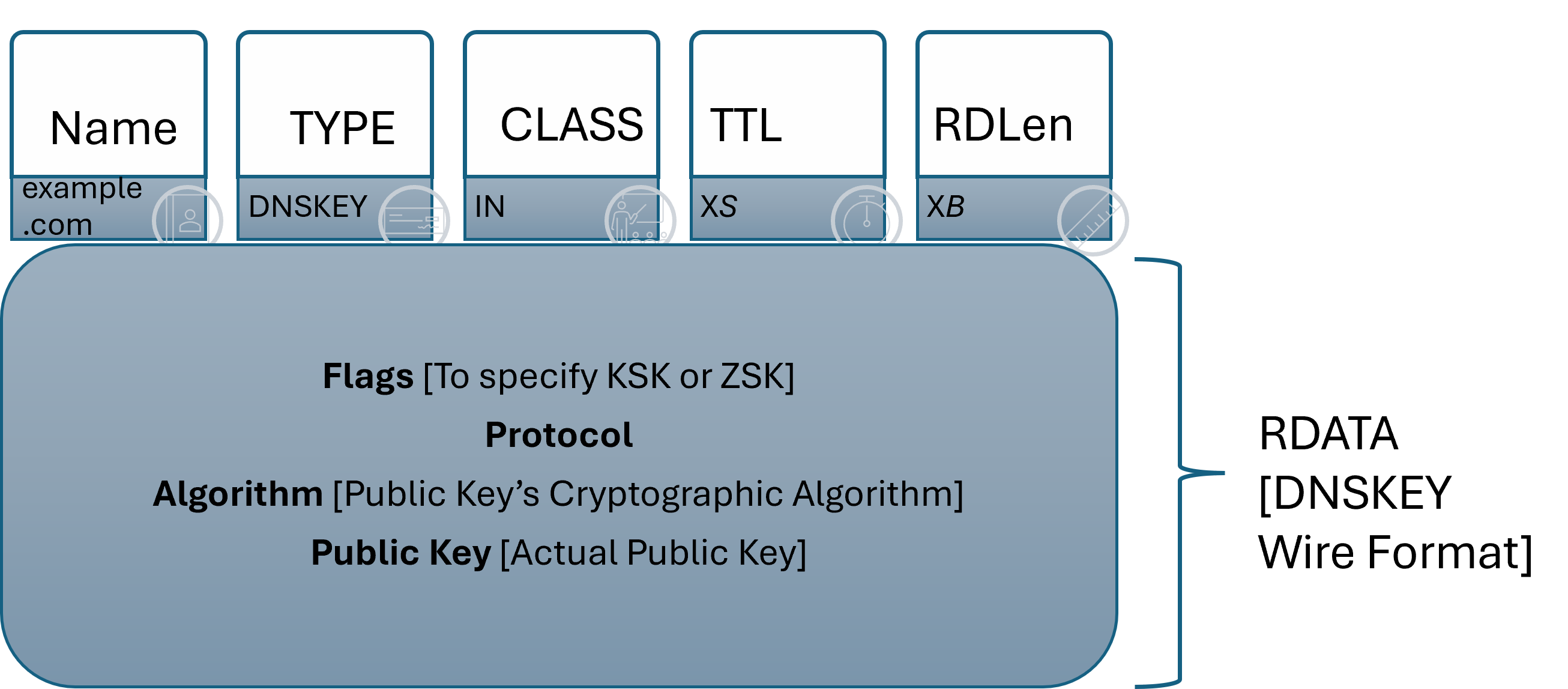}
\caption{DNSKEY RR Wire Format }
\label{DNSKEY_Format}
\end{figure}

Since RRSIG and DNSKEY RRs are the root cause of the maximum message size issue related to the double signatures (see discussion in the introduction), below we present an overview of RRSIG and DNSKEY RRs (see Figures \ref{RRSIG_Format}, \ref{DNSKEY_Format}). Their thorough understanding is pivotal for the simultaneous fragmentation of two signatures at the application layer. We refer the readers to RFC 4034 for a more detailed discussion on these relevant RRs \cite{RFC4034}.

\section{Quantum Threat \& Quantum-Safe DNSSEC}
\label{quantum_threat}
In this section, we present an overview of the impending quantum threat, its repercussions for DNSSEC, and the problems associated with sticking to conventional digital signatures or switching to post-quantum digital signatures in DNSSEC. Then, we present a discussion on our proposed quantum-safe DNSSEC.

\subsection{Quantum Threat}
The assurances of source authenticity and data integrity in DNSSEC are underpinned by digital signatures. However, these conventional digital signatures used in DNSSEC rely on the complexity and difficulty of computational problems (e.g., integer factorization and discrete logarithms), which are difficult for attackers to solve using today's classical computers \cite{BIKE_Review, Weak_key}. There exists strong evidence that these ``computationally hard problems" will become solvable in ``polynomial time" with the quantum computational model using the so-called CRQCs \cite{Mosca, CRQC}. These CRQCs will facilitate the implementation of Shor's Algorithm that will help solve the aforementioned computational problems in polynomial time \cite{shor1994algorithms}, thereby rendering the conventional digital signatures insecure. For DNSSEC, it means that in the post-quantum era (i.e., when CRQCs will be fully realized), it would be difficult to establish the integrity and authenticity of all resource records exchanged as a part of the address resolution process. Therefore, in the post-quantum era, it will become impossible to trust the final IP address of a queried domain, as the conventional digital signature will no longer offer source authenticity and data integrity at an adequate level.

Given the aforementioned impending issues with conventional digital signatures (and also with existing key encapsulation mechanisms), NIST is currently undertaking a Post Quantum Cryptography (PQC) standardization project. It aims to standardize digital signature algorithms and key encapsulation mechanisms that can run on today's classical computers but can withstand the attacks conducted using the CRQCs (if they are realized in the future). In this pursuit, NIST has recently selected three post-quantum digital signature algorithms, including FALCON \cite{Falcon}, CRYSTALS-DILITHIUM \cite{Dilithium}, and SPHINCS++ \cite{SPHINCS}. To make DNSSEC (or in general any other internet protocol) quantum-safe, it must be updated with the aforementioned post-quantum digital signature algorithms and be investigated for any issues that demand protocol-level changes to DNSSEC for the post-quantum era. However, switching from conventional digital signatures to post-quantum algorithms may be precarious. The reason is that the post-quantum digital signature algorithms as selected by NIST are currently in their infancy and thus have been studied and analyzed for a shorter period (as compared with their classical counterparts). First, there could be flaws in the algorithm design itself that make them vulnerable to cryptanalytic attacks.  Second, side-channels in the different implementations could leak keys via timing attacks, power analysis attacks, and other similar attacks. Third, there could be an algorithmic breakthrough showing how the underlying problem can be solved efficiently on a CRQC and on a classical computer. Therefore, as envisioned by the European Union Agency for Cybersecurity (ENISA) in its recent report on PQC (see details in \cite{ENSIA_Report}), the cryptanalysts may have overlooked severe attacks on post-quantum candidates (digital signatures or key encapsulation mechanisms) that could be quickly conducted using today's classical computers. Given this, simply updating global networking protocols such as DNSSEC on a large scale with post-quantum digital signature algorithms may be problematic because any attacks on these post-quantum candidates revealed in the near future will render the DNSSEC-based address resolution process untrustworthy. Conversely, keeping the conventional digital signature algorithms intact in DNSSEC (and not integrating post-quantum digital signatures) can also be problematic because CRQC inception may not be publicly announced (especially when any state or a nefarious actor builds CRQC sooner than anticipated). In such a situation, the sole reliance on conventional digital signatures for DNSSEC will make data integrity and source authentication unattainable. Therefore, the quantum threat as envisioned above makes a challenging and complex situation as of today - i.e.,  either sticking only to conventional digital signatures or switching only to post-quantum digital signatures can potentially have severe security consequences. 

\subsection{Quantum-Safe DNSSEC}

\begin{figure*}[hbt!]
\centering
\includegraphics[width=1.0\textwidth]{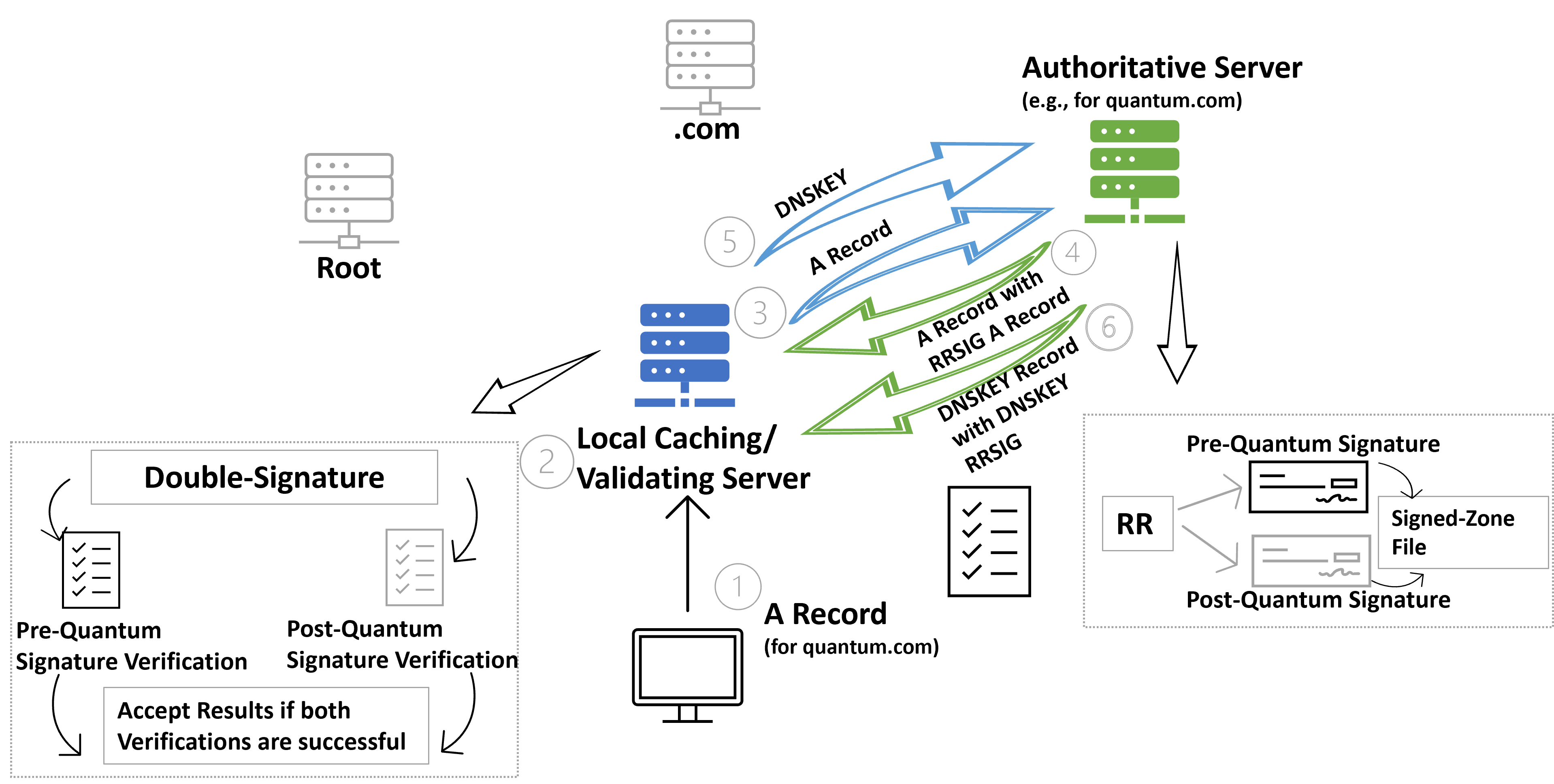}
\caption{An Overview of Double-Signed DNSSEC}
\label{fig:double-singed}
\end{figure*}

Simply replacing the conventional digital signatures with post-quantum candidates mentioned above may not make the DNSSEC quantum-safe and may even make DNSSEC susceptible to attacks conducted using today's classical computer (see the details mentioned above). Therefore, according to ENISA in \cite{ENSIA_Report}, a better and more cautious way to proceed is to use post-quantum digital signatures as an extra layer of protection in conjunction with conventional digital signatures. The combination essentially leads to a \emph{double-signed} DNSSEC (note that the terms double-signed, combiners, or hybrid cryptosystems are used interchangeably). A well-designed double-signed DNSSEC will offer the benefit that it will remain \emph{strong} if at least one of the used digital signatures is secure against a specific attack. In a double-signed DNSSEC, we will accept the IP address in a resolution process only if both signatures are verified. In particular, any attack revealed on post-quantum digital signatures in the near future that can be conducted with today's classic computer will almost always fail to compromise the address resolution process, as the security of conventional digital signatures is established and assured against classical computers. Similarly, if the post-quantum digital signatures remain secure and no attacks can be conducted upon them, they will be secure against CRQC-driven attacks. Therefore, the double-signed approach is best suited for the \emph{interim period}, as otherwise (i.e., with conventional or post-quantum digital signatures alone), DNSSEC may unknowingly be susceptible to attacks in the interim period as mentioned above.

As far as the double signing is concerned, existing literature specifies two common interfaces \cite{ENSIA_Report}. The first one, referred to as \emph{detached-message interface}, simply concatenates a pre-quantum signature \(X\) with a post-quantum signature \(Y\). The verifier only accepts the results when the \(X\) verifier accepts the \(X\) signature and the \(Y\) verifier accepts the \(Y\) signature. The second interface, referred to as \emph{signed-message interface}, typically signs with a pre-quantum algorithm to get the first signature \(X\) and then signs the result - i.e., \(X\) with the post-quantum algorithm to get the final signature (e.g, \(Z\)). The double-signature verifier in this interface normally applies the post-quantum verifier first, followed by the pre-quantum verifier (or vice versa, depending upon the signing agreement).

For the double-signed DNSSEC, we in this paper adopt an approach akin to the detached-message interface. Our reason is that this approach is relatively easier than others to integrate into our BIND9-based setup (see details on evaluation setup in Section \ref{setup}). The overview of our double-signed DNSSEC is shown in Figure.~\ref{fig:double-singed}. It essentially integrates two signatures - i.e., pre-quantum and post-quantum in the signed zone files on the authoritative server. Both these signatures are verified independently on the resolver before accepting any received resource record as verified, and 
trusted to offer enhanced security.

\section{Double-Signatures and Maximum Message Size Issue}
\label{size_issue}
In this section, we discuss the message size issue when double signatures are incorporated in DNSSEC, along with the fragmentation approach that we have devised and tested for accommodating the dual signatures in DNSSEC.
\subsection*{Messge Size Limitations}
Since DNS(SEC) is an application-layer protocol, the generated messages are first placed in UDP packets, which subsequently are placed in IP packets that make the payload of link-layer frames. The size of these payloads has a limit that is dictated by the Maximum Transmission Unit (MTU) of the physical links that these frames traverse. The size of the frame's payload is greater than the link's MTU generally enforces routers to fragment these frames into smaller IP packets that travel separately to the resolver. However, the underlying transport protocol - i.e., UDP is mainly unreliable for the successful delivery of all these fragments and thus may lead to failure of address resolution. To circumvent this issue, either we need to resort to a reliable transport protocol - i.e., TCP, or we need to formulate messages such that their size is less than the MTU of the majority of physical links over which these DNS(SEC) messages will likely travel. Since, as per our discussion in Section \ref{intro}, TCP has support issues in the existing DNS infrastructure and also is not as lightweight as UDP \cite{Frag_issue, TCP-issue, resolver_issue}, the only option is to restrict DNS message size such that IP-level fragmentation is not triggered. To accomplish this, initial RFCs on DNS \cite{RFC1035} specify the maximum DNS message size to be less than (or equal to) 512B. However, the Extension Mechanism for DNS - i.e., EDNS(0) as defined in RFC 6891 \cite{RFC6891}, theoretically allows a maximum message size of up to 64KB. Abstractly, this mechanism introduces a pseudo-RR called OPTION (alluded as OPT) and allows resolvers to specify buffer size in the `Class' field of this pseudo-RR in the `additional section' of an RR (see RR format in Fig. \ref{RR_Format}). Therefore, through EDNS(0), it is theoretically possible to increase DNS message size to a maximum of \emph{X}B if the MTU of the end-to-end path between resolver and name server is at least `X+8(UDP Header)+40(IPv6 Header). However, there is no standard way to estimate the MTU of end-to-end links. Therefore, taking the cautious approach, the maximum DNS message size is limited to 1232B so as to avoid IP-level fragmentation and frequent TCP re-transmission \cite{BIND2} (assuming MTU of 1280B).

The above-mentioned limit on the maximum DNS message size (i.e., 1232B) creates a particular challenge for double-signed DNSSEC as envisioned in this work. Precisely, any combination of pre-quantum and post-quantum signatures in double-signed DNSSEC will lead to response messages of size significantly larger than the limit discussed above. For example, our investigation shows that the message in response to a query of `Type A' (see address resolution process depicted in Figure \ref{fig:double-singed}) is generally the smallest in size. However, our thorough analysis suggests that double-signed DNSSEC with pre- and post-quantum algorithms that lead to the signatures of the smallest sizes (e.g., pre-quantum ECDSA of 64B and post-quantum FALCON512 of 690B), has an approximate response to this query of size 2500B (note that exact size depends upon variable fields in an RR). The other messages, such as those that carry `DNSKEY RRs', are even larger. Therefore, to accommodate double signatures in DNSSEC, we need to fragment these large responses at the application layer - i.e., within the DNS(SEC) protocol itself.

\subsection*{Fragmentation at the Application Layer}
There exist ways to fragment DNSSEC messages at the application layer in the post-quantum era. For example, ARRF (A Resource Record Fragment) mechanism \cite{goertzen2022postquantum} proposes a \emph{request-based} fragmentation approach at the application layer. For response messages larger than the threshold limit, ARRF on name servers `truncate' the first fragments, indicating to the resolver that there are additional fragments before this response can be re-assembled and considered complete. To facilitate the resolver in efficient reception of fragments, ARRF introduces a new pseudo-RR called Resource Record Fragments (RRFRAGS, akin to OPT described above). This pseudo-RR adopts the format of a typical RR as shown in Figure \ref{RR_Format}; however, its fields are customized such that the name server can indicate to the resolver in the first fragment how the complete response would look like once all the fragments are received. Likewise, it also facilitates name servers in ascertaining whether the resolver requests a fragment (or is it a normal query). Although ARRF seems to perform fragmentation reliably, it has two main shortcomings. The first one is that it relies on a non-standard RR that will potentially cause issues with the middleboxes. Additionally, as acknowledged by the authors of ARRF, this approach may also be susceptible to memory exhaustion attacks. 

\begin{figure*}[t!]
\centering
\includegraphics[width=0.77\textwidth]{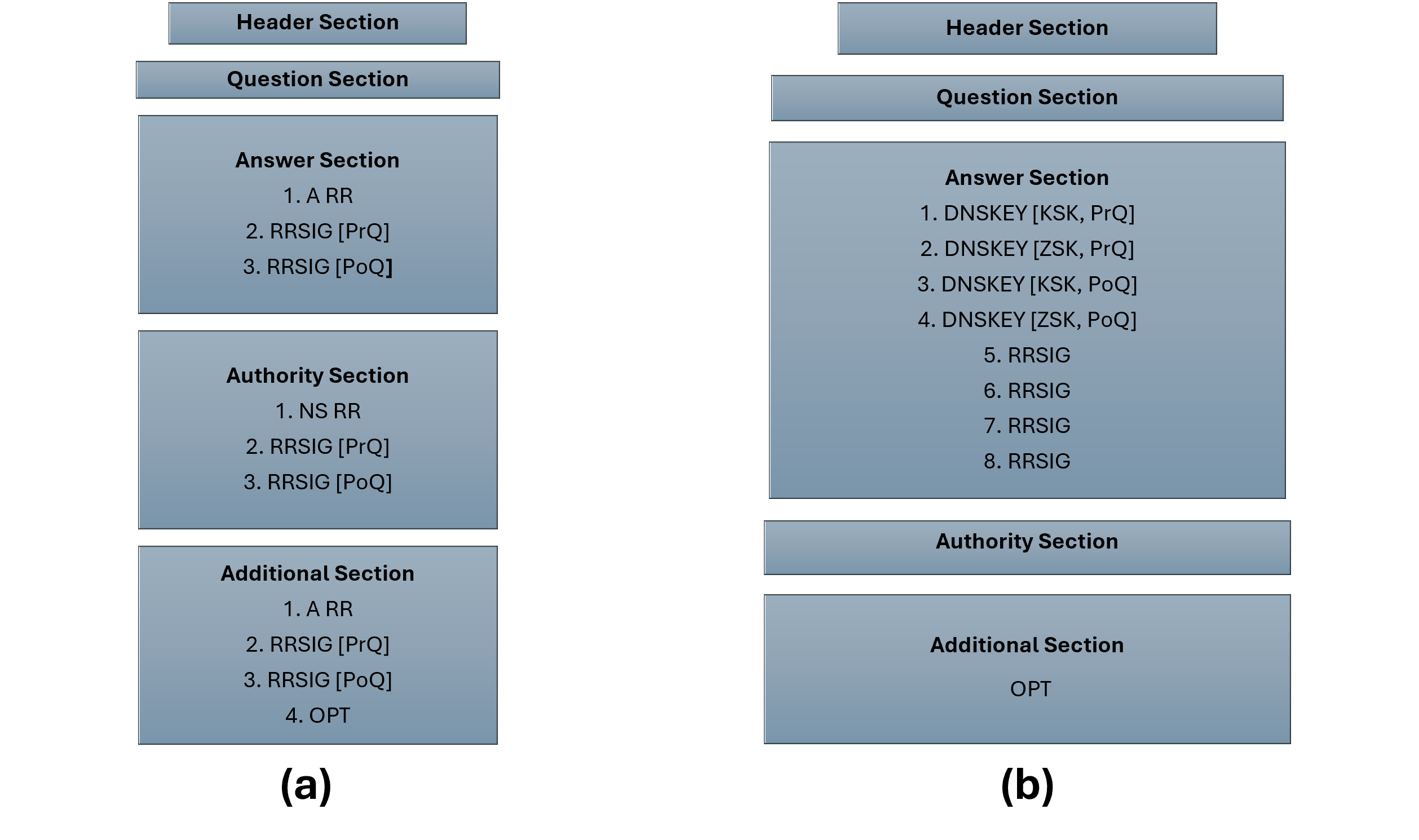}
\caption{Structure of (a) `A' DNSSEC Response  (b) DNSKEY DNSSEC Response (PrQ - Pre quantum, PoQ - Post quantum)}
\label{A_DNSKEY}
\end{figure*}
Given the above-mentioned problems with ARRF, QName-Based Fragmentation (QBF) is proposed in \cite{frag_new}. Like ARRF, QBF also adopts a request-based fragmentation approach but leverages standard RRs, unlike ARRF, which relies on customized RRs for its operations. When the response size is bigger than the threshold, QBF sends a fraction of the signature or public key in the first fragment and sets the `TC' flag in the header of the response to `1', indicating to the resolver that this response is part of a fragmented message. Using the first fragment, the resolver deduces the size of the actual response and makes explicit requests (e.g., in parallel) for all other fragments (which can be deduced from the first fragment). To indicate to the name server that a particular query is for a fragment of a previous query, the resolver makes subsequent queries as `?n?domain\_name', where `?' is a delimiter (a non-valid character for a domain, indicating that query or response is a fragment), `n' is the number of the fragment, and `domain\_name' is the domain name of the actual query. Once the `n' received by the resolver gets equal to the total number of fragments deduced by the resolver from the first fragment, all the fragments are re-assembled into the complete response and then verified.

\subsection*{Fragmentation with Double-Signatures}
The issue with both aforementioned approaches is that none of them can accommodate double signatures as envisioned in this work for the security benefits discussed in Section \ref{quantum_threat}. To accommodate two signatures while performing fragmentation at the application layer, herein, we devise an approach akin to QBF since it adopts standard RRs as discussed above. Fragmentation with double signatures poses a challenge because using a combination of pre- and post-quantum signatures and public keys results in 'A' and 'DNSKEY' responses that make it difficult for the name servers to create a first fragment with sufficient information. This first fragment is needed by resolvers to determine how many more fragments to expect before the response can be reassembled and verified.

\begin{figure}[!h]
\centering
\includegraphics[width=0.48\textwidth]{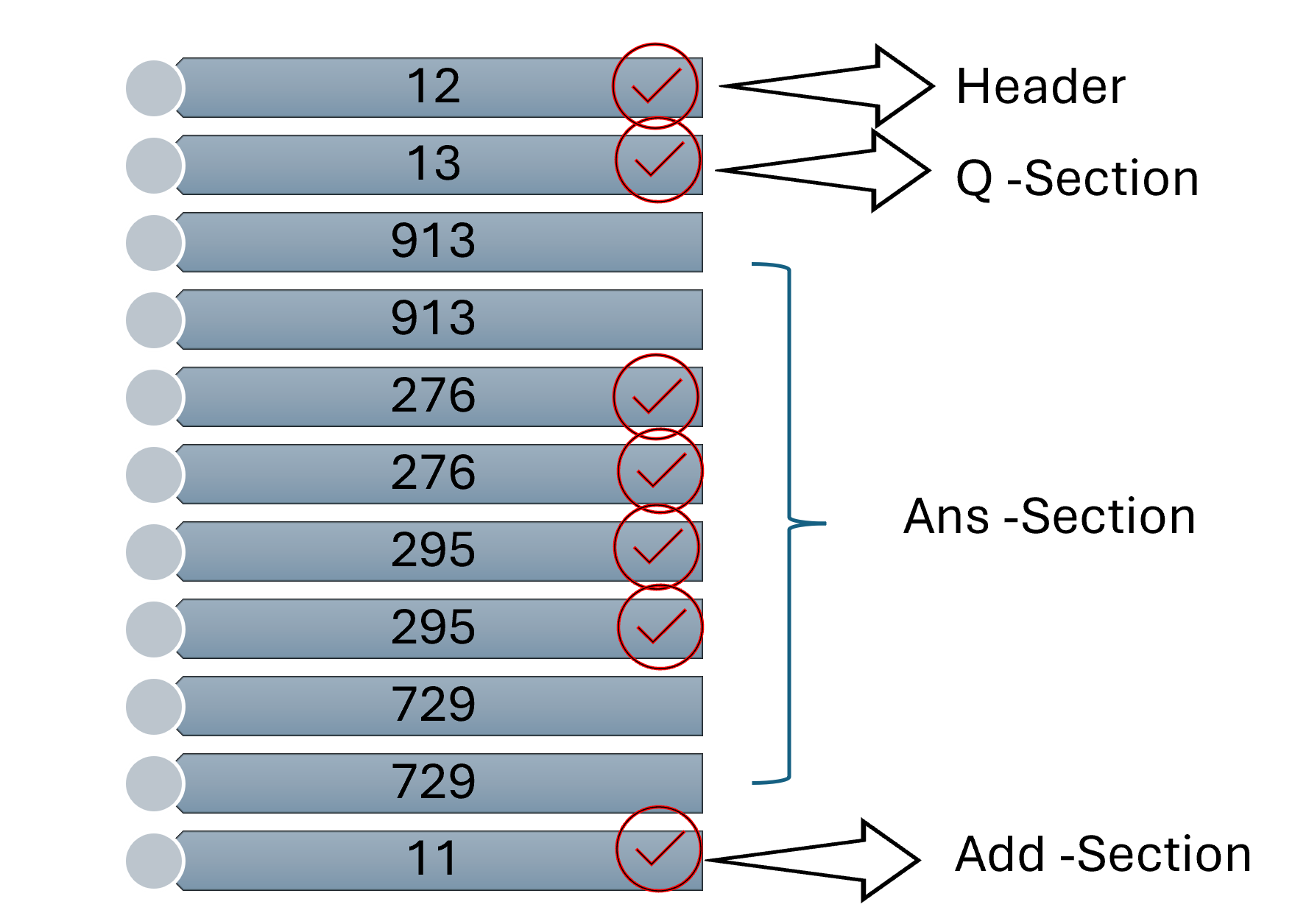}
\caption{Sizes of different fields in a representative DNSKEY Response}
\label{DNSKEY_SIZE}
\end{figure}

As per our analysis, the typical responses involved in address resolution with double signatures look something like those shown in Figure \ref{A_DNSKEY}. Since the sizes of pre-quantum signatures and public keys (e.g., for ECDSA256, RSASHA256) are much shorter than post-quantum candidates (i.e., FALCON, DILITHIUM, SPHINCS+), it is inefficient to fragment both types of RRSIGs and DNSKEYs included in typical DNSSEC responses. Therefore, we ensure that entire pre-quantum RRSIGs and DNSKEYs are always sent in the first fragment. However, our observation is that this approach works well when we are dealing with the responses to `QTYPE' of `A', leaving sufficient space in the first fragment to include a fraction of post-quantum RRSIGs. The inclusion of the entire pre-quantum and a fraction of post-quantum RRSIG in the first fragment helps the resolver in estimating the number of subsequent fragments and making explicit requests for them in `parallel' for efficiency reasons. However, our thorough analysis of the sizes of different response messages that typically make part of the address resolution process suggests that this fragmentation approach does not bode well with DNSKEY responses in certain instances (such as when RSASHA256 is used as a pre-quantum signature in double-signed DNSSEC). For example, the size of each field of all sections of a DNSKEY response to a query for `test.example' is shown in Figure \ref{DNSKEY_SIZE}. It can be observed that the total size of this response is 4462B. Note that, this response is when RSASHA256 is combined with FALCON512 - i.e., the algorithm with the smallest post-quantum signatures. The other combinations will lead to even larger response sizes. In this (and other similar) case(s), sending the pre-quantum RRSIGs and DNSKEYs along with the header, question section, and additional section (see marked fields in Figure \ref{DNSKEY_SIZE}) in the first fragment takes up around 1178B. This leaves only 54B for sending any post-quantum RRSIGs and DNSKEYs, which is insufficient. In an instance like this, it becomes entirely difficult for a resolver to estimate the remaining number of fragments and thus can not make explicit requests for them.

To circumvent the aforementioned issue, we utilize z-bits (as specified in the original RFC 1035) in the header of a DNSSEC response. Note that since two of three z-bits are used in RFC2535, however, since we adopt a daemon to accurately fragment and reassemble packets, we can still use these bits for fragmentation and reassembly in the daemon (see details of our implementation in Section \ref{setup}). Whenever a name server is unable to send any part of post-quantum signatures or keys in the first fragment (e.g., with RSASHA256), it (i.e., daemon) sets z-bits in the header (of the first fragment) such that they represent one out of three post-quantum signature algorithms (i.e., FALCON, DILITHIUM, SPHINCS+). Upon receiving a response (i.e., first fragment), the resolver daemon first checks whether z-bits are set or not. If z-bits are set, it first deduces the post-quantum algorithm and then, by leveraging pre-quantum RRs that it has already received, estimates the total size of the response and the number of fragments this response should be divided into. Subsequently, the resolver makes parallel requests explicitly for all remaining fragments.

Once all the fragments are received, the resolver first creates an updated message with both pre- and post-quantum RRSIGs and DNSKEYs and then combines all other fragments into this updated message with header values exactly as in the original message (which can be taken from the last fragment, as in the first fragment, we repurposed the z-bits). Our thorough analysis suggests that the aforementioned approach works accurately when various combinations of double signatures are incorporated in DNSSEC. However, we noticed that the re-assembled messages were not verified on the resolver (based on OQS-BIND). Our in-depth investigation suggested that this is because DNS(SEC) uses a compression scheme to reduce the size of messages by eliminating the repetition of domain names in response messages. This is accomplished by replacing the domain names with a pointer to a previous occurrence of the same domain name (see RFC 1035 \cite{RFC1035}). Due to this, our observation is that when we always send pre-quantum RRSIGs and DNSKEYs in the first fragment, it sometimes changes the order of RRs in the reassembled message (as compared with the original message created by the name server). Therefore, this issue leads to a bad compression pointer, due to which the responses fail to verify on the resolver.

As a way of dealing with this issue, we use an offset to TTLs of RRs (see Fig. \ref{RR_Format}) in DNSKEY responses to keep track of the position of pre-quantum RRs in the original response messages on name servers. Once all the fragments are received by the resolver, it leverages this offset to create an updated message that places all the RRs in their original position. Our analysis suggests that this offset approach accurately reassembles the messages such that they are correctly verified on the resolver. 

\begin{figure*}[t!]
\centering
\includegraphics[width=0.99\textwidth]{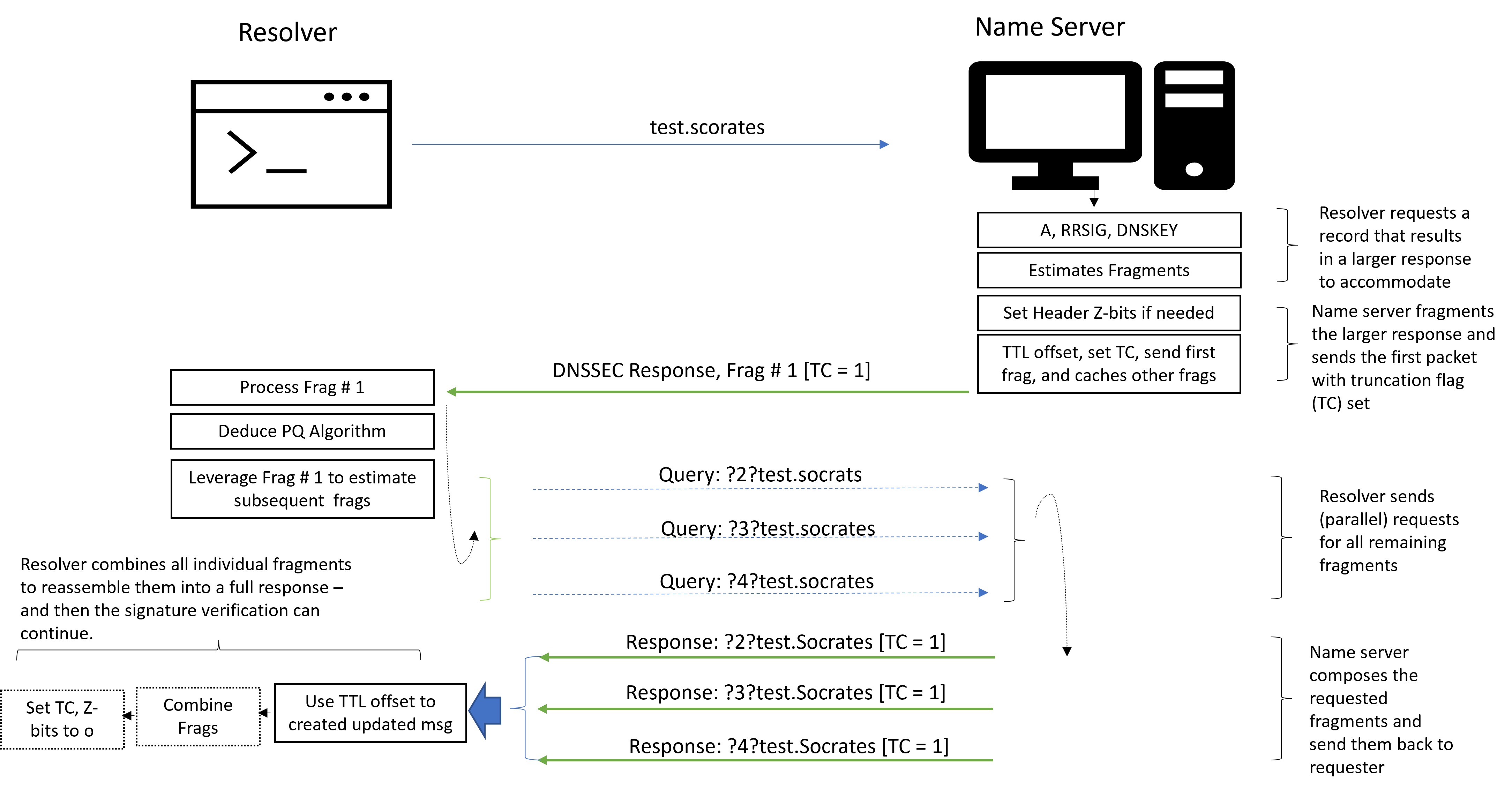}
\caption{Work-Flow of Double-Signed Fragmented DNSSEC}
\label{Overall_flow}
\end{figure*}

The overall working of double-signed fragmented DNSSEC is shown in Figure \ref{Overall_flow}.  As discussed above, whenever the response size is greater than the advertised threshold, the daemon on name server first estimates how many fragments this particular response should be divided into, sets the z-bits if needed, and includes an offset to TTL of relevant RRs, and finally sets the TC flag to `1' to indicate to the resolver that this message is fragmented. The nameserver then sends the first fragment to the resolver and caches all other fragments. When the resolver receives this first fragment, it extracts the relevant information from this fragment to estimate how many additional fragments it should expect before the original message can be reassembled. Once the fragments are estimated, the resolver then creates a separate query for all remaining fragments and sends these queries in parallel for efficiency reasons. For example, in Figure \ref{Overall_flow}, we have assumed that the original response was divided into four fragments and then the resolver makes explicit queries in parallel for fragments 2 - 4 (i.e., ?2?test.socrates - ?4?test.socrates). Note that these fragments are formatted in the same way as in \cite{frag_new}. When the name server receives these fragment requests, it sends the fragments that it has previously cached to the resolver. Once the resolver receives all the fragments (i.e., 4 in this representative example), it starts creating an updated message by utilizing TTL offset and reassembles all the fragments. Finally, it resets TC and makes the other header values as in the original message (which can be copied from the last fragments). Once done, the resolver then attempts to verify the signatures to ensure the authenticity and integrity of the received IP addresses. 

\section{Evaluation Setup}
\label{setup}
To achieve the aforementioned goals - i.e., implementing and evaluating \emph{Double-Signed Fragmented DNSSEC}, herein we present a Docker-based DNSSEC setup built on an Amazon EC2 instance. Our setup is comprised of a client, resolver, root server, and authoritative servers, all of which are running in separate Docker containers. All these components - i.e., Resolver, Root Server, and Authoritative server leverage BIND9 as a DNS software tool that is widely used in the industry. These integrated components represent the following elements of a typical DNSSEC infrastructure.

\begin{itemize}
    \item Client --- this imitates any real-world user trying to resolve a domain name and access it.
    \item Resolver --- this imitates any real-world cache server that will communicate with the DNS hierarchical servers (e.g., root server, authoritative server) on behalf of the client and facilitate the domain resolution process. Note that, since the process of caching and recursion are intimately connected, the terms recursive server and caching server are used synonymously in the literature.
    \item Root Server --- this imitates a real-world root server but contains the zone file that will only contain the NS record of our (docker-based) authoritative server.
    \item Authoritative Server --- this imitates a real-world authoritative server that contains a zone file that holds the `A' record of a dummy domain  (e.g., test.socratescrc, that we keep in zone files of our authoritative server for experimental purposes).
\end{itemize}

For experimental purposes, our `client' is a Ubuntu OS (22.04) that we simply use for querying our resolver and for analyzing the DNS response message. All of the other components - i.e., `Resolver', `Root Server', and `Authoritative' servers run instances of Internet System Consortium's BIND9 as the DNS software (see details of BIND9 below). We create a simple `Docker Network' (using Docker Version 25.0.3) to run all of the components as mentioned earlier in a separate Docker container on an Amazon EC2 t2.xlarge instance, which provides 4 cores of 2.3GHz Intel Xeon CPU and 16GB of RAM. Since by default, BIND9 does not support NIST's quantum-safe digital signature algorithms, we adopt the prototype implementation of OQS-BIND from \cite{OQS-bind} that adds the support of FALCON512, DILITHIUM2-AES, and SPHINCS+-SHA256-128S (with parameters set for level 1 - i.e, 128-bit-security) using OpenSSL 1.1.1, the Open Quantum Safe (OQS) \textit{liboqs} library 0.7.2 \cite{oqs}.

\begin{figure*}[!ht]
\centering
\includegraphics[width=1.0\textwidth]{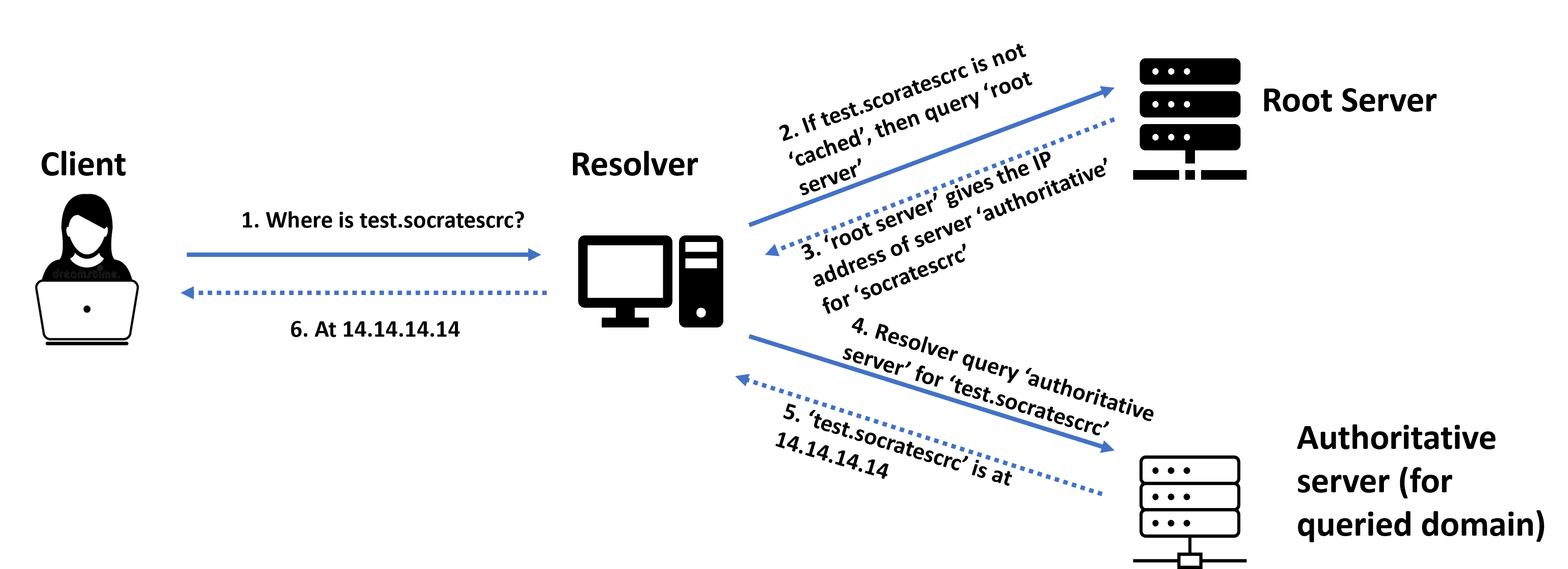}
\caption{Step-by-Step Interaction of Testbed Components}
\label{resolution}
\end{figure*}

The overall interaction of all these components as a representative DNSSEC infrastructure is depicted in Figure.~\ref{resolution}. A step-by-step description of communication among these components for a successful resolution of a domain name by our DNS(SEC) setup is presented below.

\begin{figure}[!ht]
\centering
\includegraphics[width=0.5\textwidth]{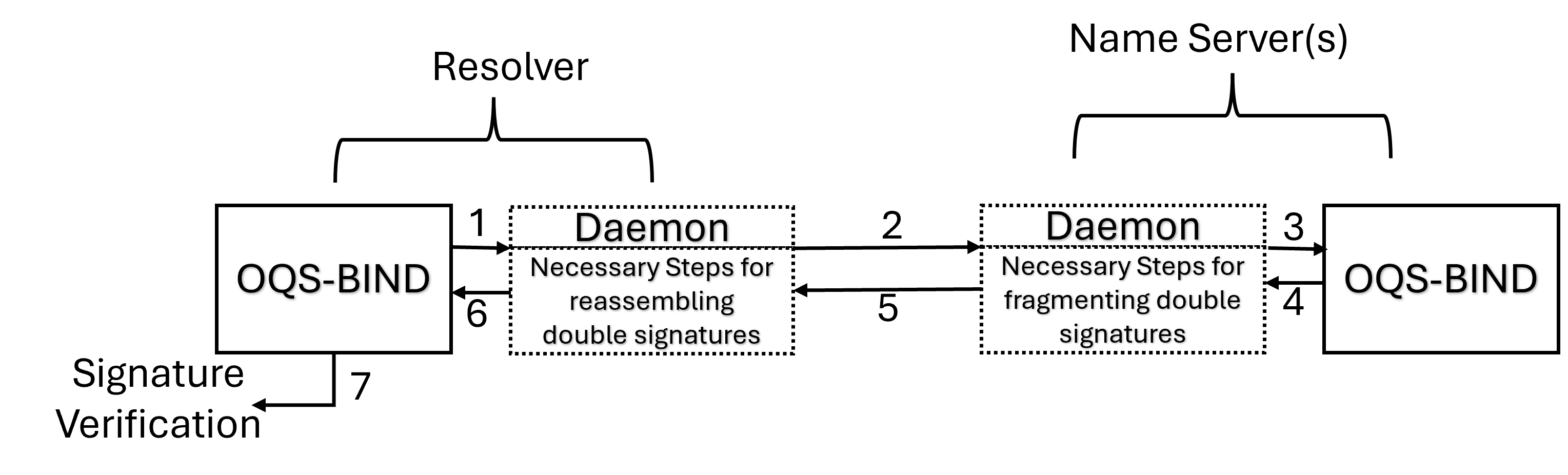}
\caption{Interaction of components of evaluation setup}
\label{interaction}
\end{figure}

\begin{itemize}
    \item[] \textbf{Step 1} Our client (a simple Docker container with Ubuntu OS) wants to resolve `test.socratescrc' (note that, for experimental purpose we have assigned an `A' RR to this subdomain on our authoritative server). It queries the `Resolver' for address resolution.
    \item[] \textbf{Step 2} Once the `Resolver' (a separate Docker container that uses BIND9) receives this query and checks whether this resource record is already cached. If yes,  it simply responds to the client with the corresponding IP (along with the other DNSSEC RRs). If not, it queries the `Root Server' to find the IP address of the server that is authoritative for the queried domain.
    \item[] \textbf{Step 3} The `Root Server' (a separate Docker container that uses BIND9) responds with the IP address of the corresponding `Authoritative Server' that can resolve the initial query.
    \item[] \textbf{Step 4} The `Resolver' queries the `Authoritative Server' whose IP address is returned by the `Root Server' (e.g., for `test.socratescrc').
    \item[] \textbf{Step 5} The `Authoritative Server' (a separate Docker container that uses BIND9) responds with the corresponding IP address along with other RRs such as RRSIG and DNSKEY needed for establishing the chain of trust.
    \item[] \textbf{Step 6} The `Resolver' validates the received response by verifying the RRSIGs and responds to the client with appropriate DNS flags set to indicate whether this response is validated by the resolver and can be trusted.
\end{itemize}
To incorporate the fragmentation strategy depicted in Figure \ref{Overall_flow} in this evaluation setup, we utilized the QBF daemon since it adopts the standard RR format, unlike ARRF that relies on custom-built RRs \cite{frag_new}. We included all the steps needed for accommodating double signatures in the daemon. This daemon then intercepts all the incoming and outgoing DNS packets on the resolver, root server, and name server and performs the relevant steps as needed (see details in Section \ref{size_issue} and Fig \ref{Overall_flow}). Therefore, practically, each component - i.e., resolver, root server, and authoritative server in our evaluation setup has an OQS-BIND software and daemon that incorporates our fragmentation strategy for double signatures (see overview in Figure \ref{interaction}). On the name servers, the daemon intercepts the incoming query, updates the maximum advertised UDP size to 64KB (i.e., max possible), and forwards it to the OQS-BIND software. The OQS-BIND creates a DNSSEC response to this query with double signatures, which is again intercepted by our daemon that executes all the steps depicted in Fig. \ref{Overall_flow}. Upon receiving the response from the name server, the resolver's daemon intercepts it and goes through all the steps needed for accommodating and re-assembling double signatures (see Section \ref{size_issue} and Fig. \ref{Overall_flow}). Once the resolver's daemon receives all the fragments, it reassembles them and finally sends the reassembled message to the OQS-BIND, which initiates the verification process for both signatures.
\subsubsection*{Berkeley Internet Name Domain (BIND)}
BIND offers a complete suite of the DNS protocol, and the BIND9 software distribution contains a name server (\texttt{named}) and a set of associated tools such as for key and signature generation, that are needed for our analysis. It is one of the most widely used software on DNS servers on the Internet \cite{BIND-Survey}. Since BIND9 can be configured (using \emph{named.conf}) as a resolver or name server, it allowed us to run `Resolver', `Root Server', and `Authoritative Server' in separate containers using the same BIND9 docker image. Although BIND9 allows flexible configuration features and supports multiple functions such as a single DNS server acting as both an authoritative name server and a resolver, we in our testbed have assigned one dedicated container to the resolver and authoritative name server as typically preferred by large operators \cite{BIND2}. Therefore, our testbed represents a realistic small-scale DNS infrastructure for assessing the impact of quantum-safe algorithms.

\subsubsection*{Double-Signed DNSSEC Setup} 
\begin{figure}[h!]
\centering
\fbox{\includegraphics[width=0.45\textwidth]{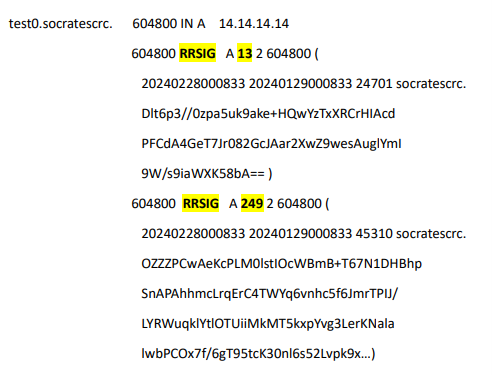}}
\caption{A portion of Signed Zone File depicting the addition of Double Signatures (Note, PQ signature - i.e., with ID 249 is shortened for brevity reasons)}
\label{Sign_SS}
\end{figure}

Our evaluation setup, as described above, is capable of performing the address resolution steps of conventional DNSSEC with both pre-quantum and post-quantum digital signatures. However, since we want the support of double signatures added to our testbed, we conducted experiments to ascertain whether BIND9 by default supports two signatures. Our observations are as follows:

\emph{Signature Generation on Name Servers:} Our analysis revealed that we can create two key pairs and signatures - i.e., both pre-quantum and post-quantum on name servers (i.e., root and authoritative servers) by using the BIND9 utilities (see \texttt{dnssec-keygen} and \texttt{dnssec-signzone} tools in the BIND9 manuals). Both signatures can be added directly to the corresponding signed zone files generated in BIND9. Figure.~\ref{Sign_SS} depicts a portion of the signed zone file generated on the authoritative server. It is evident that the signed zone file contains two signatures for a representative record -i.e., \emph{test0.socratescrc.}.

\begin{figure}[h!]
\centering
\fbox{\includegraphics[width=0.45\textwidth]{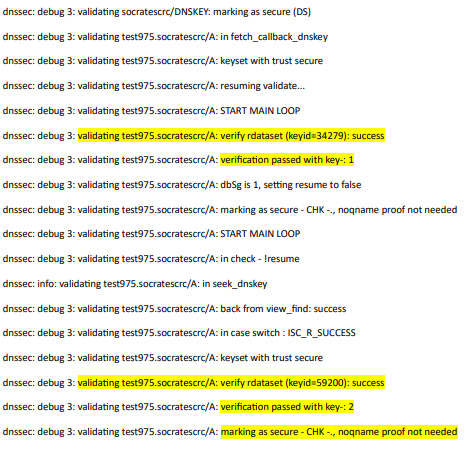}}
\caption{A portion of the `log-file' of our modified resolver that supports double-signatures}
\label{Verify_SS}
\end{figure}

\emph{Double Signature Verification on Resolver:} Our analysis of the verification process revealed that although we can send both signatures to the resolver and reassemble them correctly in our test setup, the BIND9 signature validation by default checks only one signature before marking any response from the name server(s) as secure. Therefore, we modified the BIND9 source code (especially the code corresponding to the validation of the received responses) so that both the re-assembled signatures must be valid before accepting any response. Fig.~\ref{Verify_SS} depicts a portion of the log file of our modified resolver. It is evident that the modified resolver in our setup verifies both signatures, unlike the default BIND9 resolver.

\subsubsection*{Experimental Methodology}
For analyzing the impact of double signatures in the DNSSEC address resolution process, we kept 10 A resource records with unique labels on our authoritative server (e.g., test0.socratescrc -- test10.socratescrc). Then, using the BIND9 key and signature generation tools, we generated two signatures with various possible combinations of pre-quantum (ECDSA256, RSASHA256) and post-quantum (FALCON512, DILITHIUM2, SPHINCS+-SHA256-128S) digital signature algorithms. Once both signatures were generated, we moved the DS RR of the authoritative server to the root server and the DS RR of the root server to the resolver as a trust anchor so that we could establish the chain of trust identical to the conventional DNSSEC. Once all of this process is complete, using our client, we query the resolver iteratively for 10 resource records that we kept on our authoritative server in our test setup. For each query, we estimate the query resolution time under simulated network conditions of bandwidth of 50Mbps and delay of 10ms (note that we use the Linux \texttt{tc} utility for simulating these bandwidth and latency values). Finally, we compare the average resolution time of double signatures with post-quantum signatures to find any issues in migrating from only post-quantum signatures to double signatures for the interim period until CRQCs are fully realized as discussed in Section \ref{intro}. A similar experimental approach is also used in \cite{frag_new}, but with only one signature for the post-quantum era. Additionally, we also investigate how many additional fragments our daemon needs to compute and reassemble with double signatures in comparison with post-quantum signatures alone.
\section{Experimental Results}
\label{result}

\begin{table*}[!ht]
\centering
\caption{Total Number of Fragments with Various Combinations of Double-Signatures}
\begin{tabular}{ |l|c|c| } 
 \hline
 \textbf{Digital Signature} & \textbf{Fragments in Respons to QTYPE A} & \textbf{Fragments in Respons to QTYPE DNSKEY}  \\ 
 \hline
 FALCON  & 2 & 3 \\ 
 \hline
FALCON+ECDSA  & 3 & 4\\  
 \hline
FALCON+RSA  & 3 & 4\\ 
 \hline
 DILITHIUM  & 7 & 7\\ 
 \hline
 DILITHIUM+ECDSA  & 8 & 8\\ 
 \hline
 DILITHIUM+RSA  & 8 & 8\\
 \hline
 SPHINCS  & 23 & 15\\ 
 \hline
 SPHINCS+ECDSA  & 23 & 15\\ 
 \hline
 SPHINCS+RSA  & 23 & 15\\

 \hline
\end{tabular}
\label{num_frags}
\end{table*}
In this section, we present the results of our experiments described in the previous section. Our thorough analysis suggests that our approach successfully incorporates all possible combinations of double signatures (pre-quantum + post-quantum) in DNSSEC, fragments them appropriately on name servers, and reassembles the responses correctly on the resolver side. Additionally, we also observed that all the double signatures were correctly verified on the resolver. Table \ref{num_frags} shows the number of fragments when various combinations of pre- and post-quantum signatures are employed in our DNSSEC setup. Analysis suggests that our approach only adds one additional fragment in response to QTYPE `A' and `DNSKEY' queries when FALCON and DILITHIUM2 are combined with any of the pre-quantum signatures - i.e., ECDSA256 and RSASHA256 in comparison with post-quantum signatures alone.
Furthermore, we observed that combining SPHINCS+ with any of the pre-quantum signatures does not add additional fragments. This is due to the reason that after sending the initial fragment, the daemon on name servers removes all RRs except RRSIG, DNSKEY, and OPT, thereby accommodating double-signatures created with SPHINCS+ in the same number of fragments as with SPHINCS+ alone. Mean resolution times with various combinations of double-signatures are shown in Tables \ref{num_frags_falcon}, \ref{num_frags_dil}, \ref{num_frags_sph}. The results show that the impact of double signatures is insignificant on the resolution times of queries when compared with post-quantum signatures alone. In any double-signature combination that we have tested, resolution time only increases by less than 9\% as compared with the post-quantum signatures alone. For instance, the resolution time for a combination of FALCON + ECDSA is 205.9ms as compared with 190.1ms for FALCON alone (i.e., an increase of approximately 8.3\%). The lowest resolution time for double signatures was observed for any combination of FALCON with either RSA or ECDSA-based pre-quantum digital signatures. This is because FALCON has the shortest signature size compared with other post-quantum digital signatures. Similarly, RSA combined with any of the three post-quantum candidates results in a lower resolution time as compared with ECDSA (e.g., DILITHIUM + RSA results in a lower resolution time in comparison with DILITHIUM + ECDSA, and likewise for other two post-quantum digital signatures). This is because RSA has faster signature verification than ECDSA \cite{Signing_Speed}.

The aforementioned observations suggest that double signatures can reliably be incorporated into DNNSEC infrastructure and have a minimal impact on performance as compared with post-quantum digital signatures alone. Therefore, their incorporation is ideally suited for DNSSEC during the interim period from now until CRQCs are fully realized. Particularly, the double signatures created with either FALCON or DILITHIUM with RSA or ECDSA have a resolution time that is 14-22\% less than that of SPHINCS+ alone. Since both FALCON and DILITHIUM are based upon lattices, SPHINCS+ was only added as a post-quantum candidate by NIST to avoid the over-reliance on the lattices (see \cite{NIST2}). However, as observed in our experiments, SPHINCS+ has a significantly larger signature size, which results in much higher resolution time in DNSSEC. Therefore, the double-signature created with either FALCON or DILITHIUM is the optimal solution for the specified interim period while avoiding the envisaged problems of SPHINCS+ (see results described above and depicted in Tables \ref{num_frags_falcon} - \ref{num_frags_sph}). These double-signatures will essentially offer additional security to DNSSEC during the interim period, if any attacks on lattice-based signatures are revealed in the near future that could be perpetrated through today's classical computer.

\begin{table}[!ht]
\centering
\caption{Means Resolution Time (ms) - Various Combinations of FALCON512, ECDSA256, and RSASHA256}
\begin{tabular}{|p{1.6cm}|p{2.8cm}|p{2.95cm}|} 
 \hline
 \textbf{FALCON} & \textbf{FALCON+ECDSA} & \textbf{FALCON+RSASHA}  \\ 
 \hline
 190.1  & 205.9 & 204.5 \\ 
 \hline

\end{tabular}
\label{num_frags_falcon}
\end{table}

\begin{table}[!ht]
\centering
\caption{Means Resolution Time (ms) - Various Combinations of DILITHIUM2, ECDSA256, and RSASHA256}
\begin{tabular}{|p{1.6cm}|p{2.8cm}|p{2.95cm}|} 
 \hline
 \textbf{DILITHIUM} & \textbf{DILITHIUM+ECDSA} & \textbf{DILITHIUM+RSASHA}  \\ 
 \hline
 214.5  & 225.6 & 220.7 \\ 
 \hline

\end{tabular}
\label{num_frags_dil}
\end{table}

\begin{table}[!ht]
\centering
\caption{Means Resolution Time (ms) - Various Combinations of SPHINCS+-SHA256-128S, ECDSA256, and RSASHA256}
\begin{tabular}{|p{1.6cm}|p{2.8cm}|p{2.95cm}|} 
 \hline
 \textbf{SPHINCS+} & \textbf{SPHINCS+ECDSA} & \textbf{SPHINCS+RSASHA}  \\ 
 \hline
 245.6 & 263.3 & 256.7 \\ 
 \hline

\end{tabular}
\label{num_frags_sph}
\end{table}

\section{Related Work}
\label{related_work}
Our literature review suggests that, historically, there have been a few proposals for dealing with large message size issues. For example, authors in \cite{hist_frag1, hist_frag2} proposed potential ways for conducting fragmentation at the application layer. However, both of them can potentially lead to DoS amplification attacks since they send many packets in response to a single query. Additionally, there are a few recent research works that focus on transitioning toward quantum-safe DNSSEC and related issues. However, unlike this work, none of these prior research works consider double-signatures for DNSSEC. For example, authors in \cite{Muller, Beernink,goertzen2022postquantum} investigated NIST Round 3 and 4 PQC candidates, respectively, for their utilization in DNSSEC. Müller et al.~\cite{Muller} specifically analyzed NIST Round 3 candidates and established the various requirements that different candidates either meet or do not for their incorporation in DNSSEC. Similarly, Beernink et al.~\cite{Beernink} presented a way of delivering larger keys out-of-band from DNSSEC. Goertzen et al.~\cite{goertzen2022postquantum} focused on the fragmentation issue in DNSSEC for the post-quantum era. However, as discussed before, it adopts a custom-designed non-standard pseudo-RR for performing the fragmentation at the application layer. The usage of non-standard RR makes its practicability questionable. Similarly, an improved version of the fragmentation approach for quantum-safe DNSSEC is also presented in \cite{frag_new}, which leverages only standard RRs and thus seems more practical. However, unlike these existing works, we investigate the possibility of using double signatures for DNSSEC, and how the fragmentation issue can be dealt with in the simultaneous presence of pre- and post-quantum signatures in DNSSEC. Through our implemented testbed and empirical analysis, we show the efficacy of double signatures in DNSSEC for the interim period until CRQCs are fully realized. 

\section{Conclusion \& Future Work}
\label{conclusion}
In this work, we investigated the possibility of using double signatures in DNSSEC. For this purpose, we first investigated the maximum message size issue resulting from the incorporation of double signatures in depth and then devised an approach that can successfully incorporate double signatures on name servers and deliver them to the resolver over UDP by conducting the fragmentation and re-assembly of response messages at the application layer. We then built a functional Docker-based testbed by leveraging OQS-BIND9 and a daemon that handles double signatures, fragmentation of relevant response messages, and the reassembly of the received fragments. Our investigation suggests double signatures can be generated and added to signed zone files on the authoritative name servers, and our resolver can facilitate the verification of two signatures for the same resource record. With these required changes made to the BIND9 resolver and clubbing our daemon for fragmentation/re-assembly of double-signatures with it, we comprehensively analyzed the resolution time for various combinations of pre- and post-quantum signatures. Our analysis revealed that double signatures do not have any significant overhead as compared with post-quantum digital signatures alone. Particularly, our investigation suggests that lattice-based signatures - i.e., FALCON and DILITHIUM combined with any of the pre-quantum digital signatures result in a shorter resolution time than SPHINCS+ alone, which was selected (despite having a very large signature size) by NIST to avoid over-reliance on the lattice-based signatures for the post-quantum era. Our analysis suggests that double signatures generated with FALCON and DILITHUM are well-suited for DNSSEC and offer performance benefits compared with SPHINCS+ alone. The double signatures approach will also offer additional security in case any attacks on lattice-based PQ candidates are revealed in the near future. While the analysis conducted in this work establishes the overall efficacy of double signatures for DNSSEC, we believe that resolution time can further be reduced by incorporating our fragmentation strategy directly into BIND9 rather than in the form of a daemon. We plan to make these advancements and relevant analyses in the future. 


\section*{Acknowledgment}
\label{Acknowledgments}

\noindent This work was supported by the Cyber Security Cooperative Research Centre Limited, whose activities are partially funded by the Australian Government's Cooperative Research Centres Program. 

\bibliographystyle{ACM-Reference-Format}
\bibliography{main}


\end{document}